\documentclass[10pt,journal,twocolumn,twoside]{IEEEtran}
\usepackage{amsfonts}
\usepackage{amssymb,amsmath,mathrsfs}
\usepackage{bm} 
\usepackage{mathtools}
\mathtoolsset{showonlyrefs}
\usepackage{cite}
\usepackage[pdftex]{graphicx}
\usepackage{epstopdf}
\usepackage{float} 
\usepackage{relsize}
\newtheorem{example}{Example}

\usepackage{flushend}
\usepackage[dvipsnames]{xcolor}

\definecolor{forestgreen}{rgb}{0.0, 0.27, 0.13}

\ifCLASSOPTIONcompsoc
\usepackage[caption=false,font=normalsize,labelfont=sf,textfont=sf]{subfig}
\else
\usepackage[caption=false,font=footnotesize]{subfig}
\fi
\title{On the Error Floor Evaluation of \\ NOMA-Irregular Repetition Slotted ALOHA}
\author{Estefan\'ia Recayte, \IEEEmembership{Member, IEEE }
\thanks{Estefan\'ia Recayte is with the Institute of Communications and Navigation of the German Aerospace Center (DLR), 82234 Wessling, Germany. Email:\{\texttt{estefania.recayte}\}\texttt{@dlr.de}.  }
\thanks{Copyright (c) 2025 IEEE. Personal use of this material is permitted. However, permission to use this material for any other purposes must be obtained from the IEEE by sending a request to pubs-permissions@ieee.org.
}

}

\newcommand{\users}{\ensuremath{{m}}}
\newcommand{\usersBalls}{{\bar{m}}}
\newcommand{\slots}{{n}}
\newcommand{\load}{\mathsf{G}}

\newcommand{\bins}{{b}}

\newcommand{\plr}{\mathtt{PLR}}

\usepackage{acro}
\DeclareAcronym{AWGN}{short = AWGN ,long = additive white Gaussian noise}
\DeclareAcronym{ACRDA}{short = ACRDA ,long = asynchronous contention resolution diversity ALOHA}
\DeclareAcronym{CDF}{short = CDF ,long = cumulative distribution function}
\DeclareAcronym{CRA-CC}{short = CRA-CC ,long = CRA-convolutional code}
\DeclareAcronym{CRA-SH}{short = CRA-SH ,long = CRA-shannon bound}
\DeclareAcronym{CRA}{short = CRA ,long = contention resolution ALOHA}
\DeclareAcronym{CRDSA}{short = CRDSA ,long = contention resolution diversity slotted ALOHA}
\DeclareAcronym{CRDSA++}{short = CRDSA++ ,long = contention resolution diversity slotted ALOHA++}
\DeclareAcronym{CRI}{short = CRI ,long = contention resolution interval}
\DeclareAcronym{CSA}{short = CSA ,long = coded slotted ALOHA}
\DeclareAcronym{CSI}{short = CSI ,long = channel state information}
\DeclareAcronym{DAMA}{short = DAMA ,long = demand assigned multiple access}
\DeclareAcronym{DSA}{short = DSA ,long = diversity slotted ALOHA}
\DeclareAcronym{DSSS}{short = DSSS ,long = direct sequence spread spectrum}
\DeclareAcronym{E_SSA}{short = E-SSA ,long = enhanced spread spectrum ALOHA}
\DeclareAcronym{ECRA}{short = ECRA ,long = enhanced contention resolution ALOHA}
\DeclareAcronym{ECRA-SC}{short = ECRA-SC ,long = ECRA selection combining}
\DeclareAcronym{ECRA-MRC}{short = ECRA-MRC ,long = ECRA maximal-ratio combining}
\DeclareAcronym{EGC}{short = EGC ,long = equal-gain combining}
\DeclareAcronym{FEC}{short = FEC ,long = forward error correction}
\DeclareAcronym{GEO}{short = GEO ,long = geostationary orbit}
\DeclareAcronym{HAP}{short = HAP ,long = high-altitude platform}
\DeclareAcronym{IC}{short = IC ,long = interference cancellation}
\DeclareAcronym{IoT}{short = IoT ,long = Internet of things}
\DeclareAcronym{IRA}{short = IRA ,long = irregular repetition ALOHA}
\DeclareAcronym{IRCRA}{short = IRCRA ,long = irregular repetition contention resolution ALOHA}
\DeclareAcronym{IRSA}{short = IRSA ,long = irregular repetition slotted ALOHA}
\DeclareAcronym{LDPC}{short = LDPC ,long = low-density parity-check}
\DeclareAcronym{LEO}{short = LEO ,long = low-Earth orbit}
\DeclareAcronym{M2M}{short = M2M ,long = machine-to-machine}
\DeclareAcronym{MAC}{short = MAC ,long = medium access control}
\DeclareAcronym{MF}{short = MF ,long = matched filter}
\DeclareAcronym{MF-TDMA}{short = MF-TDMA ,long = multi-frequency time division multiple access}
\DeclareAcronym{MRC}{short = MRC ,long = maximal-ratio combining}
\DeclareAcronym{MUD}{short = MUD ,long = multiuser detection}
\DeclareAcronym{pmf}{short = pmf ,long = probability mass function}

\DeclareAcronym{PDF}{short = PDF ,long = probability density function}
\DeclareAcronym{PER}{short = PER ,long = packet error rate}
\DeclareAcronym{PLR}{short = PLR ,long = packet loss rate}
\DeclareAcronym{QPSK}{short = QPSK ,long = quadrature phase-shift keying}
\DeclareAcronym{RA}{short = RA ,long = random access}
\DeclareAcronym{RCB}{short = RCB ,long = random coding bound}
\DeclareAcronym{RTT}{short = RTT ,long = round trip time}
\DeclareAcronym{SA}{short = SA , long = slotted ALOHA}
\DeclareAcronym{SB}{short = SB ,long = Shannon bound}
\DeclareAcronym{SC}{short = SC ,long = selection combining}
\DeclareAcronym{SIC}{short = SIC ,long = successive interference cancellation}
\DeclareAcronym{SNIR}{short = SNIR ,long = signal-to-noise and interference ratio}
\DeclareAcronym{SINR}{short = SINR ,long = signal-to-interference and noise ratio}
\DeclareAcronym{SNR}{short = SNR ,long = signal-to-noise ratio}
\DeclareAcronym{TDMA}{short = TDMA ,long = time division multiple access}
\DeclareAcronym{UCP}{short = $\Code$-UCP ,long = $\Code$-unresolvable collision pattern}
\DeclareAcronym{VF}{short = VF ,long = virtual frame}

\DeclareAcronym{rv}{short = r.v., long = random variable}

\usepackage{subfig}
\IEEEoverridecommandlockouts

\begin{document}

\maketitle

\thispagestyle{empty} \setcounter{page}{0}

\begin{abstract}
In this work, we provide a simple yet tight analytical approximation of the  packet loss rate in the error floor region for a non-orthogonal multiple access (NOMA)-based  \ac{IRSA} scheme. Considering an Internet of Things (IoT) scenario, users randomly select both the number of replicas based on a designed degree distribution and the transmission power from predetermined levels, while successive interference cancellation (SIC) is performed at the receiver.  Our derived packet loss rate expression in the finite length regime is promptly evaluated. 
{Its accuracy is validated through Monte-Carlo simulations, demonstrating a strong match across channel loads, including those beyond the low load regime.}
\end{abstract}

\begin{IEEEkeywords}
Non-orthogonal multiple access, irregular repetition slotted ALOHA, error floor evaluation, finite length 
\end{IEEEkeywords}

\section{Introduction}

\begin{figure*}[t!]
  \centering
  \subfloat[$\mathcal{S}_1$, $\mu_1 = 2$ slots, $\nu_1 = 2$ users]{
  \includegraphics[width=.3\textwidth]{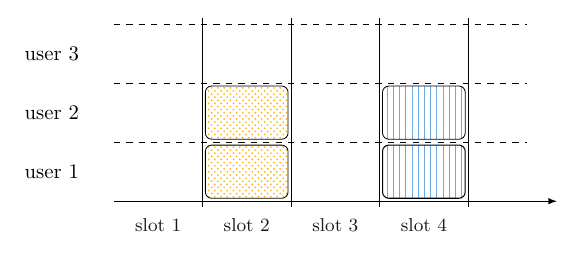}
  \label{fig:plrDiffDistrib}
  }\hspace{1em}
  \subfloat[$\mathcal{S}_2$, $\mu_2 = 3$ slots, $\nu_2 = 3$ users]{
  \includegraphics[width=.3\textwidth]{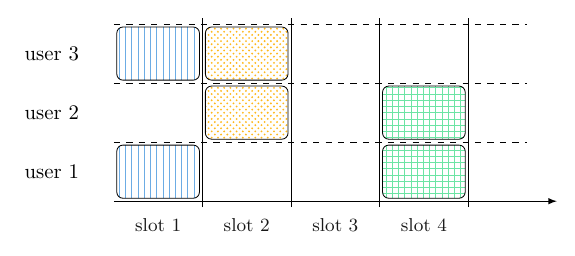}
  \label{fig:truDiffDistrib}
  }\hspace{1em}
  \subfloat[$\mathcal{S}_3$, $\mu_3 = 3$ slots, $\nu_3 = 2$ users]{
  \includegraphics[width=.3\textwidth]{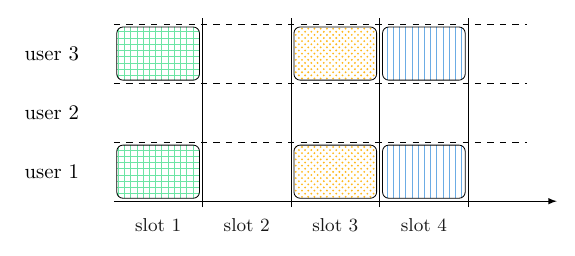}
  \label{fig:truDiffDistrib}
  }
  \caption{Stopping sets in NOMA-based IRSA with the corresponding number of users and slots. The filling color-patterns represent different power levels. }
  \label{fig:stoppingsets}
\end{figure*} 
The convergence of the Internet of Things (IoT), massive-type machine communications (mMTC), and 6G technology promises to revolutionize connectivity, enabling a vast array of low-cost devices to seamlessly exchange data \cite{IoTintro}.   A crucial aspect of IoT and mMTC is implementing efficient medium access methods capable of handling sporadic transmissions from a massive number of  devices. \emph{Modern random access} protocols play a pivotal role in this context \cite{Berioli16}, offering flexible and scalable solutions to manage connectivity among the dynamic and often unpredictable nature of IoT deployments.  These protocols enable devices to access the network without prior coordination, optimizing resource utilization while ensuring reliable and timely data transmission. 
 
  A notable contribution in modern random access schemes was provided in  \cite{Liva:IRSA}, with the development of  irregular repetition slotted ALOHA (IRSA). This protocol  combines  packet diversity with \ac{SIC}, enabling the resolution of collisions. In IRSA, users transmit a variable number of replicas over a frame according to a predesigned distribution.  The IRSA protocol was further investigated by incorporating power diversity per user in \cite{MengaliCRDSA, Hmedoush:multi} and per  replica  in \cite{CRDSApowerDiv, RecayteIRSA}. The advantage of power diversity lies in exploiting the capture effect, which enables the SIC algorithm to solve collisions even  when multiple users transmit 
   in the same slot. 

  In parallel to this, NOMA-based schemes,  originally proposed for downlink transmissions, were also introduced for random access in the pioneering work  \cite{ChoiNoma}. NOMA schemes rely on accurately defining a set of power levels from which users can choose. This approach ensures that collisions involving packets with different power levels can be resolved effectively. Thus, the SIC algorithm benefits from both the diversity of transmitting multiple packets and the diversity of power levels.
  In this context, NOMA-based random access schemes have been the subject of recent research.

Bounds on the throughput of the NOMA-ALOHA scheme were presented in \cite{NOMA:fan}, and the optimal load  was analyzed in massive IoT scenarios with sparse user activity. 
{In \cite{Babich:energy} the throughput and a bound of the success probability for a NOMA scheme both for CRDSA with two transmissions and IRSA in fading channel was evaluated.} 
In \cite{Huang:iterative},  authors investigate the asymptotic performance of a slotted ALOHA scheme with NOMA to predict throughput performance.  The analysis assumes that users can place their replicas within an infinitely long frame. This assumption allows for the effective use of density evolution  to understand the system's behavior.   Similarly, asymptotic analysis using density evolution was also conducted for a special case of IRSA, namely \ac{CRDSA}, in  \cite{Rama:CRDSA}. 
  In \cite{Su:NOMA}, the replica degree distribution and power distribution are jointly optimized in the asymptotic regime using density evolution  for NOMA-coded slotted ALOHA. This modern random access   protocol   integrates packet-level erasure correcting codes with SIC techniques. Similarly, \cite{Noma:shao} presents an asymptotic analysis for NOMA-IRSA, focusing on the optimization of degree distribution and power levels.

  In the existing literature, most studies consider the performance of NOMA  random access schemes under the assumption of infinitely long frames \cite{NOMA:fan, Huang:iterative, Rama:CRDSA, Su:NOMA, Noma:shao}, { primarily evaluating  system behavior at high channel loads. However, to the best of our knowledge,  their performance in the non-asymptotic regime is unexplored. To address this gap and complementing the existing  literature, this work investigates the performance of NOMA-IRSA schemes in the more practically relevant case of finite frame lengths.} Specifically, we derive a tight analytical approximation of the packet loss rate for low and moderate channel loads, i.e., in the error floor region.
  
Unlike the accurate yet involved approximation  proposed in \cite{AlexEF2} and \cite{shortframe},  we present here a computationally efficient one-shot expression derived from the balls in bins (BiB) problem. In \cite{shortframe}  authors present an approximation for the IRSA scheme under the assumption of very short frames, limited to a maximum of 50 slots. In contrast, this work evaluates the NOMA-IRSA scheme, demonstrating its validity even for significantly longer frames. {Furthermore, we propose a simplified method for evaluating the error floor, offering  a compact, analytically tractable form that requires significantly fewer operations compared to \cite{AlexEF2, EF_Alex, Enrico_EF}. } 

This  work also provides an extension of the study presented in \cite{Hmedoush}, which estimates the packet loss rate in a multipacket detection scenario  under conditions of low channel loads and for a specific degree distribution.  In the results section, we show that the solutions proposed in this work provide more accurate results for the NOMA-IRSA scheme compared to the evaluation given in \cite{Hmedoush}.

  Finally, we validate the proposed expression through Monte-Carlo simulations, showing a strong match even for systems operating at moderate channel loads.

  The rest of the paper is organized as follows. Section~\ref{sec:sysmodel} presents the system model, while in Section \ref{sec:plr} the packet loss rate approximation is derived. Numerical results are given in Section~\ref{sec:numres}, and Section~\ref{sec:con} concludes.

\section{System Model}\label{sec:sysmodel}

We consider a system where $\users$ uncoordinated IoT active nodes (users) transmit packets over a shared uplink wireless channel to a common receiver. A \ac{MAC} frame is composed by $\slots$ consecutive slots where the duration of a slot equals a packet duration. Users are frame and slot synchronized.
 The system channel load $\load$ is  \[{\load }= \frac{\users}{\slots} \quad \text{[packets/slot]}. \]
A NOMA-\ac{IRSA} scheme is considered. Following the \ac{IRSA} protocol, each user sends $r$ replicas of the generated data packet  and uniformly places them across the  $\slots$ available slots in a frame. The number of replicas $r$ to transmit  is determined by a predefined degree distribution. The polynomial formulation of IRSA degree distribution can be written as \cite{Liva:IRSA}
\begin{equation}\label{eq:irsadist}
    \Lambda(x) = \sum_{r=2}^{r_{\max}} \lambda_r x^r,  \text{ with }
\sum_{r=2}^{r_{\max}} \lambda_r = 1,
\end{equation}
where  $\lambda_r$ represents the probability that a user transmits $r$ replicas.   
Each replica has a pointer indicating where its copies are placed over the frame.  This information is used by the receiver during the decoding process.

After determining the number of replicas to transmit, the user selects the transmission power level for each replica uniformly from the $L$ available options, as suggested in \cite{ChoiNoma}. We denote the power levels by
\begin{equation}
 p_1 > ... > p_L > 0, \text{ with }   p_k = \gamma (I_k + 1),
\end{equation}
where $\gamma$ is the target \ac{SINR} and $I_k = \sum_{j = k+1}^{L} p_j$  with $I_L = 0$.  With easy manipulations,   the $k$-th power level  can be written as 
\begin{equation}\label{eq:power}
    p_k = \gamma (\gamma + 1 )^{L-k}. 
\end{equation}
 
At the receiver side, SIC is performed. Assuming the use of capacity achieving codes \cite{Liva:IRSA}, the receiver can decode a replica whenever its SINR, denoted by $\Gamma$,   meets or exceeds a predefined threshold, i.e,  $\Gamma \geq \gamma$.
Note that the definition of the power levels ensures that interference-free replicas can always be decoded. Additionally, a slot containing replicas transmitted at different power levels can be resolved by iteratively cancelling the highest-power replica from the slot.
Once a replica is decoded, the receiver can identify the positions of its copies and subtract their contribution from the frame \cite{Liva:IRSA}.  The decoder may iterate multiple times over each slot and the entire frame until all replicas are decoded or no more replicas with $\Gamma > \gamma$  are present. 

{Let us explain the behavior of the SIC algorithm with a simple example, focusing on the case where multiple replicas are received in the same slot.}
\begin{example} \emph{
{Let us consider an example to illustrate a case where decoding fails.
 Suppose there are $L=3$ power levels, and two users transmit in the same slot using the same power level, e.g. $p_2 =  \gamma (\gamma +1)$.
To successfully decode a replica, its SINR, must be at least 
$\gamma$. If two replicas are transmitted in the same slot with the same power level, the SINR for either replica is given by
\[\Gamma = \frac{\gamma (\gamma +1)}{1 + \gamma (\gamma +1)}.\]
This value is strictly less than $\gamma$ and the receiver fails in such a scenario.  Conversely, let us consider when three users transmit their replicas in the same slot, but at different power levels.
\\
The power values are $p_1 = \gamma (\gamma +1 )^2, p_2 =  \gamma (\gamma +1 ), p_3 = \gamma$. The SINR of the user who transmitted at $p_3$ is 
\[
\Gamma = \frac{\gamma(\gamma + 1)^2}{1 + \gamma(\gamma + 1) + \gamma} = \gamma.
\]
Since this meets the decoding threshold \( \gamma \), the replica with  transmitted power $p_3$ is successfully decoded. Its contribution is then canceled via SIC and the SINR of the user who transmitted at $p_2$ becomes
\[\Gamma = \frac{\gamma(\gamma + 1)}{1 + \gamma} = \gamma
\]
which again is sufficient for decoding. After this second replica is cancelled, the remaining one is interference-free and is decoded with an SNR of \( \gamma \). \\ }}
\end{example}

\section{Packet Loss Rate Approximation}\label{sec:plr}
\begin{figure*}[t!]
\setcounter{equation}{4}
 \begin{align}
 \begin{split}
  \plr  \approx
  \frac{1}{L^2 }\frac{2 \, (\lambda_2)^2 \, \users}{\slots (\slots -1 ) }  +  
    \frac{1}{L^3} \;  \frac{2 (\slots-2 ) (\lambda_2 \, \users  )^2 }{ \slots^2 (\slots-1)^2 } 
    + \frac{1}{L^3} \,
    \frac{ 6 \, (\lambda_3)^2 \, \users}{ \slots \, (\slots-1) \, (\slots-2) }. 
   \end{split}
 \label{eq:plrfinal}
 \end{align}
 \hrulefill
 \end{figure*}

In this section, we consider results from the balls into bin (BiB) problem to derive a tight one-shot formula approximation of the packet loss rate, denoted by $\plr$.

\subsection{Balls into bin  problem}
The general BiB problem, see e.g. \cite{Johnsonbook}, consists in independently throwing $\usersBalls $ balls into $\bar{\bins} $ bins. This setup can be mapped to the NOMA-IRSA scenario, where each ball represents a user transmitting its replicas, and $\bar{\bins}$ bins represent all possible combination of  slots in which a user can place the replicas. 

Following this parallel, multiple balls landing in the same bin correspond to multiple users transmitting their replicas in the same slots. 
The BiB problem allows us to evaluate distribution of the number of bins containing $t$ balls. Let $Y_t$ be the random variable (rv) denoting the number of bins containing exactly $t$ balls. Then, from \cite{Johnsonbook}, we have that
\begin{equation}\label{pr_y}\setcounter{equation}{2}
\Pr\{Y_t = y_t \} = \frac{\usersBalls!}{\bar{\bins}^{\bar{\bins}}} \sum_{k=0}^{\bar{\bins} -  y_t} (-1)^k \binom{y_t + k}{y_t}\frac{(\usersBalls - k)^{\usersBalls  - y_t}}{( t!)^k (\usersBalls  - k t)!}.
\end{equation}
As shown in \cite{Johnsonbook}, for $\usersBalls  \rightarrow \infty $  the rv $Y_t$ can be approximated to a Poisson random variable with parameter
 $\beta_t$, i.e. 
 \begin{equation}\label{eq:pois}
  {Y_t \sim \mathcal{P}(\beta_t)} \text{ and }  \beta_t = \frac{\bar{\bins}}{t!} \Big(\frac{\usersBalls}{ \bar{\bins}}\Big)^t.  \,  
 \end{equation}

\subsection{Packet loss rate as BiB}\label{sub:plr}
The packet loss rate, $\plr$, is defined as the probability that a user is not successfully decoded at the end of the SIC process. In IRSA, the $\plr$ in the error floor region can be approximated by analyzing   the so call \emph{stopping sets} \cite{AlexEF2}. A stopping set is a configuration where the decoder gets stuck because multiple users transmit their replicas in the same slots, preventing any of them  from being decoded. 
A stopping set is characterized by 
$\nu$, the number of users, and 
$\mu$, the number of slots involved, respectively.
{
When power diversity is introduced, as in the NOMA-IRSA scheme, the receiver employing SIC   fails to decode users in the error floor region if two independent events occur: (i) the users’ replicas are part of a stopping set, and (ii) every slot within this stopping set contains packets transmitted at the same power level.
To account for the event where replicas are transmitted with the same power, we proceed as follows. }
We denote  by  $\sigma(\mu)$ the event in which each of the $\mu$ slots of an stopping set contain colliding replicas transmitted at the same power. 
Considering in total $L$ different level of power, we have   \[ \Pr\{\sigma(\mu)\} = \frac{1}{L^{\mu}}.\] 
We evaluate the error floor  by accounting for the three most probable stopping sets that can occur for NOMA-IRSA, denoted with $\mathcal{S} = \{\mathcal{S}_1, \mathcal{S}_2, \mathcal{S}_3\}$. An example of each stopping set is illustrated in Fig.~\ref{fig:stoppingsets}.  The stopping set $\mathcal{S}_1$ consists of two users transmitting two replicas each  in the same time slots at the same power. The stopping set $\mathcal{S}_2$  consists of three slots and three users, each transmitting two replicas different time slots, resulting in collisions where each user interferes with the other two. Also in this case, each slot has colliding replicas at the same power.
 The stopping set  $\mathcal{S}_3$ consists of two users transmitting three replicas each in the same three slots at the same power.
 Note that $ \mathcal{S}_1 $ and $ \mathcal{S}_2 $ can occur only if the IRSA distribution \eqref{eq:irsadist} includes a non null probability for a user transmitting two replicas, i.e., if $\lambda_2 >0 $. Similarly, $ \mathcal{S}_3$ can only occur if $\lambda_3 > 0 $. 
 Note that these are the most representative stopping sets for the NOMA-IRSA scheme under evaluation. As demonstrated in the results, they are sufficient to provide an accurate evaluation of the performance.

 \begin{table}[t]  
  \centering
  \caption{}
  \vspace{-.5em} Parameters of considered stopping sets for $\users$ users and $\slots$ slots 
  \label{tab:par}
  \vspace{1em}
  \begin{tabular}{r|c|c|c|c} %
  $ i$      & $\mu_i$ & $\!\!\nu_i\!\!\!$ & $\usersBalls$  & $\bar{\bins} = b(\mu_i,\nu_i )$\\
    \hline
    \hline      
   $1$  & $2$ & $2$ & $\lambda_2 \, \users$ & $\!\! \binom{\slots}{2} $  \\[.1cm]
   $2$  & $3$ & $3$ & $\lambda_2 \, \users$ & $\!\!     {\binom{\slots}{2}}/{\sqrt{2 \, (\slots -2)}} $      \\[.1cm]
   $3$  & $3$ & $2$ & $\lambda_3 \, \users$ & $\!\! \binom{\slots}{3} $ 
    \end{tabular}  
\end{table}
{We derive the probability of the stopping set $\mathcal{S}_i$ occurs  by casting the scenario of users transmitting in the same slots as a BiB problem.  

 Given $\users$ users and $\slots$ slots, the probability that the stopping set $\mathcal{S}_i$ occurs $u$ times within a frame is given by ${\Pr\{Y_{\nu_i} = u }\} $. 
 {The average number of users that can participate in the stopping sets (corresponding to the number of balls) is given by $\bar{m} =\lambda_2 m$ for stopping set $\mathcal{S}_1$ and $\mathcal{S}_2$, and $\bar{m} = \lambda_3 m$ for stopping set $\mathcal{S}_3$.}
The parameter values for each stopping set, $\nu_i, \mu_i, \usersBalls$ and $\bar{\bins} = b(\mu_i,\nu_i )$,  are provided in Table~\ref{tab:par}.
 In the appendix section, we discuss and explain the derivation $b(\mu_2, \nu_2) = b(3,3)$ which does not have a straightforward correspondence to the classic balls-in-bins problem.

{The packet loss rate is approximated by computing, for each stopping set, the probability that users are part of it. This probability is calculated as the ratio of users involved in a collision (positive cases) to the total number of transmissions (total cases), and the probability that replicas within the same slot are transmitted at the same power, as follows}
\begin{align} 
\plr \,\, &  \approx \sum_{ \mathcal{S}_i \in \mathcal{S}}  \Pr\{\sigma(\mu_i)\} \frac{  \sum_{u = 1}^{\infty}\nu_{i}  u \Pr\{ Y_{\nu_{i}} = u\}}{\users}   
\\ 
&  = \sum_{ \mathcal{S}_i \in \mathcal{S}} \frac{1}{L^{\mu_i}} \frac{\nu_{i}}{\users} \sum_{u = 1}^{\infty} u \Pr\{ Y_{\nu_{i}} = u \}.
\end{align}
{ Recalling that $Y_{\nu_i}$ can be considered Poisson distributed, then $ \Pr\{ Y_{\nu_{i}} = u \} =  \frac{(\beta_{\nu_i})^{u}}{u!} \, e^{-\beta_{\nu_i}} $ we have can write the previous expression as  }  
\begin{align}
  \plr 
  &  \approx \sum_{ \mathcal{S}_i \in \mathcal{S}} \frac{1}{L^{\mu_i}} \, \frac{\nu_{i}}{\users}  \sum_{u = 1}^{\infty}  u \, \frac{(\beta_{\nu_i})^{u}}{u!} \, e^{-\beta_{\nu_i}}.
\end{align}
Observing that \[\sum_{u = 1}^{\infty}  u \, \frac{(\beta_{\nu_i})^{u}}{u!} e^{-\beta_{\nu_i}} = \beta_{\nu_i},\] we have
\begin{align}
  \plr  & \approx\sum_{ \mathcal{S}_i \in \mathcal{S}} \frac{1}{L^{\mu_i}} \frac{\nu_{i}}{\users} \, 
  \frac{\bar{\bins}}{\nu_i!} \Big(\frac{\usersBalls}{ \bar{\bins}}\Big)^{\nu_i}  \\
  & \approx\sum_{ \mathcal{S}_i \in \mathcal{S}} \frac{1}{L^{\mu_i}}  \, 
  \frac{1}{(\nu_i- 1) !} \Big(\frac{\usersBalls}{ \bar{\bins}}\Big)^{\nu_i -1 }. \label{eq:plrf}
\end{align}
Finally, by substituting  the values from Table~\ref{tab:par} into each stopping set in \eqref{eq:plrf} and performing straightforward manipulations, we derive the approximated equation for the packet loss rate given in \eqref{eq:plrfinal}.

Equation \eqref{eq:plrfinal} addresses the challenge of dealing with more complex formulations often required for evaluating the packet loss rate in the error floor region, as seen in works such as \cite{AlexEF2, EF_Alex, Enrico_EF, shortframe}. {The proposed expression reduces complexity through a compact and analytically tractable form.}

While one might initially consider the proposed expression a bound for the $\plr$, given that it accounts for a finite number of stuck occasions for the receiver, this claim cannot be made rigorously. {This limitation arises because \eqref{eq:plrfinal} is based on approximations.  First,   the number of users transmitting either two or three replicas  is estimated using the average values $\bar{\mathsf{m}} = \lambda_2 \, \mathsf{m}$ and $\bar{\mathsf{m}} = \lambda_3 \, \mathsf{m}$, respectively. Second, the number of slots containing exactly $\nu_i$ users, denoted by $Y_{\nu_i}$, is approximated by a Poisson random variable. These two approximations prevent the resulting packet loss rate expression from being formally classified as a bound.
}

\section{Numerical Results}\label{sec:numres}
\begin{figure}[t!]
    \includegraphics[width=\columnwidth]{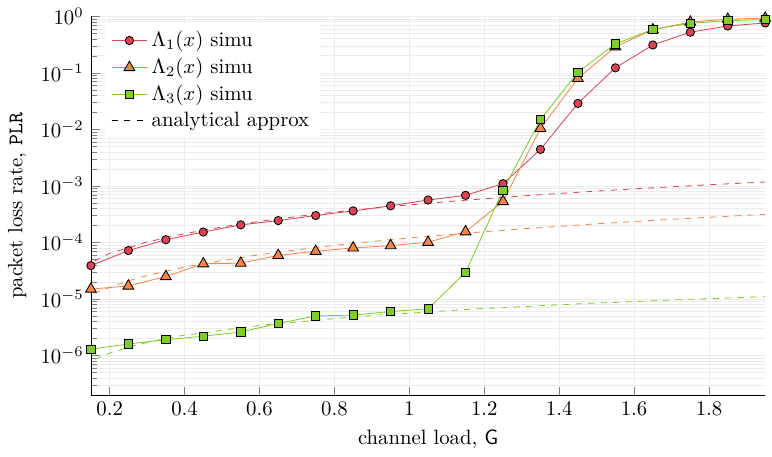}
       \centering \caption{ Packet loss rate as a function of channel load  for NOMA-IRSA for $\slots = 200$ slots,  $L=3$ power levels, and  $  \Lambda_1(x)$,
        $\Lambda_2(x) $, and $\Lambda_3(x)$. Curves with markers represent simulation results, while dashed lines correspond to the analytical approximation given in \eqref{eq:plrfinal}. }
       \label{fig:plr}
    \end{figure} 

This section examines the validity of the proposed $\plr$ approximation. We consider different setups and compare the simulation results with the analytical form presented in \eqref{eq:plrfinal}.

In our first scenario we consider a frame length of ${\slots = 200}$ slots, target SINR $\gamma = 3$dB, and  ${L= 3}$ different levels of power. According \eqref{eq:power}, we have $ {p_1 = 17.9007}$, ${ p_2 = 5.9763}$, \  and  $ {p_3 = 1.9953.}$ 
For $\slots = 200$, $\mu_2 = 3 $ and $\nu_2 = 3$, the number of bins is given by $b(3,3) = 1000$. 
The packet loss rate, $\plr$, for NOMA-IRSA, plotted as a function of the channel load, is shown in Fig.~\ref{fig:plr}. The following degree distributions, commonly employed in literature and known to offer good performance \cite{Liva:IRSA}, are considered \begin{align}
    \Lambda_1(x) & = 0.50 x^2 + 0.50 x^3, \\
    \Lambda_2(x) & = 0.25 x^2 + 0.60 x^3 + 0.15 x^8,  \text{ and   } \\
    \Lambda_3(x) & = x^3.
\end{align}

The marker curves correspond to Monte-Carlo simulations while dashed lines represent the packet loss rate approximation given in \eqref{eq:plrfinal}.  Notably, the derived $\plr$ accurately predicts the simulated behavior across all considered IRSA degree distributions in the error floor region. Note that the error floor region for NOMA-based IRSA schemes includes not only low channel loads but also moderate channel loads. 
Focusing on a $\plr$ of practical relevance, e.g. $10^{-3}$, we observe that channel load up to 1.20 [packets/slot] can be supported. 

These initial results demonstrate that the proposed expression is effective for evaluating a NOMA-based IRSA. The provided expression offers a fast and simply method for assessing the performance of the system, remaining accurate even at very low values of packet loss rate.

Note that the derived approximation is valid in the error floor region but, as expected, diverges under high channel loads, i.e., in the waterfall region. This divergence arises because the behavior of the SIC algorithm at high loads is influenced by factors beyond the stopping sets. Specifically, the SIC process can stall due to excessive traffic. An analytical evaluation in the waterfall region for NOMA-IRSA can be implemented assuming infinitely long frames, a scenario that has been extensively documented in the literature, for example, in \cite{NOMA:fan, Huang:iterative, Rama:CRDSA, Su:NOMA, Noma:shao}.

In our second scenario, represented in Fig.~\ref{fig:plr2}, we  plot the packet loss rate as a function of the channel load for degree distribution $\Lambda_1(x) = 0.50 x^2 + 0.50 x^3$. This is done for different levels of power $L$ considered, with frame length $\slots =200$ slots and target SINR of $\gamma =3$. The aim of these results is to compare our evaluation with the findings proposed in \cite{Hmedoush}. 
As expected, the packet loss rate improves as the number of power levels increases. This is because the receiver can resolve more collisions, thanks to a higher probability of encountering colliding replicas with different power levels. Note that there is a trade-off between increasing the number of power levels, which leads to higher energy consumption, and the system's performance. This balance requires a dedicated investigation.
The dashed lines represent the proposed approximation, which remains accurate in the error floor region for any number of power levels considered. Dotted lines are results where only considers the stopping set $\mathcal{S}_1$ as proposed in \cite{Hmedoush}. It can be observed that the approximation in \cite{Hmedoush}  holds well only for low channel loads and becomes less accurate as the channel load increases to moderate levels. This is because, for many degree distributions, as the number of transmitting users increases, the probability of  $\mathcal{S}_2$ and $\mathcal{S}_3$ occurring becomes non negligible. This plot demonstrates that, for any value of $L$, the proposed analytical evaluation consistently outperforms the performance of the existing approximation.

\begin{figure}[t]
    \includegraphics[width=\columnwidth]{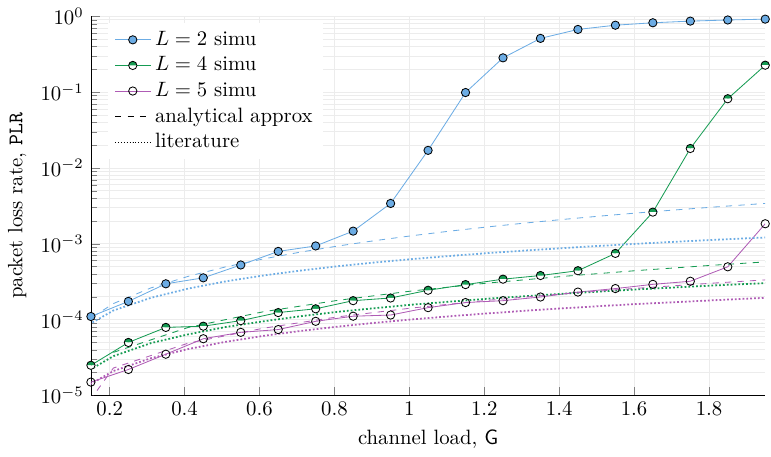}
       \centering \caption{ Packet loss rate  as a function of channel load for NOMA-IRSA with $\slots = 200$ slots,  $\Lambda_1(x)$, and different powers levels. 
      Curves with represent simulation results, dashed lines indicate the approximation given by \eqref{eq:plrfinal}, and dotted lines shows results from literature in \cite{Hmedoush}. }
       \label{fig:plr2}
    \end{figure}

    \begin{figure}[t]     
        \includegraphics[width=\columnwidth]{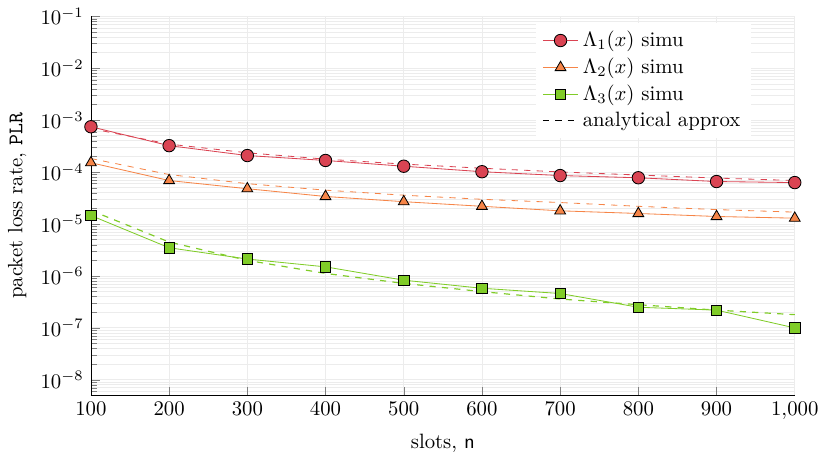}
           \centering \caption{ Packet loss rate against frame length  for NOMA-IRSA at $\load  = 0.80$ [packets/slot], $L=3$ power levels, for $  \Lambda_1(x) , \Lambda_2(x)$, and $\Lambda_3(x)$. 
          Curves with markers represent simulation results, while dashed lines correspond to the analytical approximation given in \eqref{eq:plrfinal}. }
           \label{fig:plrslots}
        \end{figure}

Finally, we further evaluate the proposed approximation for different \ac{MAC} frame lengths. 
In our last scenario we consider $\Lambda_1(x), \Lambda_2(x),$ and $\Lambda_3(x)$ for $L=3$ power levels, target SINR $\gamma = 3$, and fixed channel load $\load = 0.80$ packets/slot.
As shown in Fig.~\ref{fig:plrslots}, the approximation remains accurate for every value of $\slots$ slots considered in the frame and across all IRSA distributions. Increasing the number of available slots decreases the packet loss rate, as the probability of users experiencing lock situations becomes lower.

\section{Conclusions}\label{sec:con}
In this work, the packet loss rate for NOMA-IRSA scheme is evaluated, targeting IoT and mMTC applications with finite frame length. A tight approximation of the $\plr$ performance is presented and validated through Monte-Carlo simulations. The results demonstrate a perfect match for different degree distributions, across different power levels and frame lengths.
The proposed analysis provide valuable insights for IoT network design by offering a simple and fast computational evaluation of the system's performance.

Additionally, while previous studies focus on the the asymptotic analysis of the NOMA-IRSA scheme, this work complements them by providing an evaluation of finite-length behavior within the error floor region.
\begin{appendices}\label{appendix1}
\appendix{ }
The evaluation of number of bins $\bar{\bins} = b(\mu_i,\nu_i)$ for stopping sets $\mathcal{S}_1$ and $\mathcal{S}_2$ is straightforward,  as these cases are align with the BiB problem. Recalling that, balls represent users, while bins represent the possible combinations of slots where users can place their replicas. When two balls land in the same bin, it corresponds to users transmitting their replicas in the same slots. 
 In contrast,  the interpretation of $\bar{\bins} = b(\mu_i,\nu_i)$ for the stopping set $\mathcal{S}_3$ is not applicable because the number of slots involved in the stopping set ($\mu_i = 3$) differs from the number of slots where a user  place their replicas (two). 

To find a analog formulation to BiB for this case we empirically derive a solution by first determining the packet loss rate given by users involved only in the scenario given by stopping set $\mathcal{S}_3$ through  simulations, i.e. $\plr(\mathcal{S}_3, \slots)$. Simulations are  performed for different values of slots $\slots$ for a fixed channel load, e.g. $\load = 0.4$ [packets/slot], and one power level $L=1$.

  Two key observations are fundamental to have an evaluation of \( \bar{\bins} \). First, from equation \eqref{eq:plrf}, we can express \( \bar{\bins}^2 \) as \[ \bar{\bins}^2 = \frac{\bar{\users}^2}{2 \plr(\mathcal{S}_3, \slots)}. \] Second, simulation results revealed that the trend of $\bar{\bins}^2$ as a function of $ \slots $ closely aligns with the behavior of \( \binom{\slots}{2} \). Given this trend, we hypothesize that 
  \[ \ {\bar{\bins}^2} 
   = \binom{\slots}{2}^2 \frac{1}{g(\slots)^2},
  \]
  where \( g(x) \) is a linear function of the form \( g(x) = a_0 + a_1 x \). To proceed, we use a fitting function (e.g. in Matlab) to determine the coefficients \( a_0 \) and \( a_1 \), allowing us to approximate the relationship between \( \bar{\bins} \) and \( \slots \). As a result, we obtain the values \( a_0 = -4\) and \( a_1 = 2 \), allowing us to write
  \begin{equation}
    b(\mu_2,\nu_2) = \bar{\bins} =\binom{\slots}{2} \frac{1}{\sqrt{2(n-2)}}.
  \end{equation}
 
\end{appendices}

\section*{Acknowledgement}
Estefan\'ia Recayte acknowledges the financial support by the Federal Ministry for Research, Technology and Space (BMFTR) in Germany in the programme of “Souverän. Digital. Vernetzt.” Joint project 6G-RIC, project identification number: 16KISK0-PIN.

\flushend
\bibliographystyle{IEEEtran}
\bibliography{IEEEabrv,references}

\end{document}